# Rate-Dependent Tensile Response of Polyvinyl Chloride Geomembranes

Nesrin Akel[1*], Guillaume Stoltz[1], Antoine Wautier[1], François Nicot[3], Nathalie Touze[2]

[1]*INRAE, Aix Marseille Univ, RECOVER, Aix-en-Provence, France*

[2]*Université Paris-Saclay, INRAE, SDAR, Jouy-en-Josas, France*

[3]*Université Savoie Mont-Blanc, ISTerre, Le Bourget-du-Lac, France*

*corresponding author: nesrin.akel@inrae.fr*

## Abstract

One of the challenge that face the effectiveness of Polyvinyl Chloride geomembranes (PVC GMs) as a hydraulic barrier is the capacity to withstand unexpected mechanical actions, particularly tensile forces, during installation and throughout their lifespan. These forces pose risks of premature failure and impermeability degradation. In this study, the characterization of the short and long-term mechanical response of PVC GMs to uniaxial tensile forces has been investigated. Uniaxial tensile test have been performed for tensile rates spanning several orders of magnitude. Analysis of the true stress-strain curves reveals a significant decrease in tensile modulus, strength, and strain at failure at low strain rates, which are relatively close to those applied in situ. Long-term investigations have been conducted as well, through relaxation tests. Our key results unveil two distinct characteristic times in stress relaxation, with the fast relaxation occurring over the first four hours. During this phase, the pre-relaxation loading rate affects the relaxation behavior. Beyond this phase, the relaxation behavior becomes independent from the pre-relaxation loading rate. Burger's rheological model is proposed to measure the stress relaxation at different rates. The model's results validate the existence of two characteristic times.

# 1. Introduction

Rapid climate change, economic crises, pollution, and drought crises – all of which are the result of population increase and industrialization – highlight the urgent need for sustainable development and new solutions to planetary healing. Geosynthetics are emerging as a key tool for developing new adaptation strategies to worldwide challenges and mitigating opportunities (Dixon et al., 2017; Koerner et al., 2008; Touze, 2020).

Geomembranes, a particular type of geosynthetics, play an important function as barriers to limit fluid migration. They are contributing to sustainable water conservation strategies by preventing water loss along dams, irrigation canals and reservoirs (Cazzuffi et al., 2010; Giroud and Plusquellec, 2017; Koerner et al., 2017, 2008). Moreover, they are preferred over common barrier materials (i.e., Compacted Clay Liner CCL) for their extra advantages, which include impermeability, cost-effectiveness, ease of construction, and sustainability (Cazzuffi and Gioffrè, 2023a).

Polyvinyl Chloride (PVC) geomembranes, the focus of this study, are synthetic membranes known for their remarkable flexibility and deformability. These properties make them indispensable for waterproofing various types of both existing and newly constructed dams (Cazzuffi et al., 2012, 2010; Cazzuffi and Gioffrè, 2023b, 2023a, 2020; Giroud, 2021; Scuero and Vaschetti, 2017). Additionally, its applications extend to long-term water management in underground structures such as tunnels (Cazzuffi et al., 2012; Luciani and Peila, 2019).

Although PVC geomembranes are not designed to transmit stresses, they face challenges from mechanical actions that could compromise their integrity, and consequently lead to a loss of waterproofing property (Luciani and Peila, 2019). Those stresses may arise both during installation, in the short term, and along the life of the work in which they are installed, on the long term. Thus, the success of projects involving geomembranes relies on the appropriate design, installation and use to avoid any rupture and consequent leak (Cazzuffi et al., 2010; Luciani and Peila, 2019).

Among the configurations that have been identified to give rise to unexpected tensile forces generated inside improperly installed geomembranes is the case of tension at the bottom of a slope when the geomembrane lacks support and a load is generated by water contained in a reservoir or a dam for example. Furthermore, PVC geomembranes might experience differential settlement due to a non-uniform underlying surface with subsequent risks of failure (Bennett and Brachman, 2024; Cen et al., 2018; Eldesouky and Brachman, 2023; Fox and Thielmann, 2014; Giroud and Soderman, 1995; Marcotte and Fleming, 2021). In that case, the geomembrane interacts with underlying angular stones, which may result in excessive localized tensile stress in the geomembrane, leading to puncturing or bursting failure (Abdelaal et al., 2014; Brachman et al., 2018; Cen et al., 2021; David Frost et al., 2012; Fan and Rowe, 2023; Hornsey and Wishaw, 2012). Therefore, the distribution of these tensile forces within the geomembrane, and consequently, the ability of a geomembrane to elongate without failing is strongly important.

Several regulations and standardized tensile tests have been put in place to evaluate the mechanical properties of the geomembrane. However, one significant factor to consider is the disparity in tensile speeds between laboratory tests and in situ scenarios. The actual tensile speed is not necessarily of the same order of magnitude as in a typical laboratory test performed in normalized procedures, like the European standard EN12311 (AFNOR, 2013) that recommends a 100 mm/min displacement rate, or ASTM D4885 (ASTM D4885, 2018) that recommends a 10 mm/min displacement rate. Thus, studying the tensile speed effect seems to be of key importance for determining the mechanical response and the resistance of PVC GMs, as studied by (Merry and Bray, 1997) specifically on HDPE geomembrane.

To date, numerous studies have shown an importance of studying the environmental conditions effects like temperature and chemical effects on the mechanical response (Blanco et al., 2018; Cazzuffi and Gioffrè, 2021; Eun et al., 2018; Hsuan et al., 2008; Karademir and Frost, 2021; Stark and Choi, 2005; Xiao et al., 1997; Zhang et al., 2021). Additionally, investigations into plasticizer retention effects have been conducted (Blanco et al., 2017; Cazzuffi and Gioffrè, 2021;

Giroud and Tisinger, 1993; Hsuan et al., 2008, p. 200; T. Stark et al., 2005; Stark et al., 2004; T. D. Stark et al., 2005). However, to the best of our knowledge, it is noteworthy that a limited number of studies have placed importance on rate-dependent effects in the mechanical response of PVC geomembranes. Wu et al. (2023) have examined the effect of tensile rates on the relaxation behavior by considering three rates within the same order of magnitude (2.5, 5, and 10 mm/min). They reveal that the stress relaxation behavior of PVC GMs is considerably impacted by both the tensile rate and the initial strain. Specifically, an increase in initial strain and tensile rate results in more time for the relaxation stress to stabilize.

A major engineering challenge is to use short-term laboratory test results to design structures that will protect the environment and public health over decades. Although the stress magnitude in real structures is expected to be much lower than the material's breaking strength, unexpected local situations can cause stress concentrations. Therefore, the present paper examine the impact of these unexpected tensile stresses on the tensile properties (tensile strength, elastic modulus, elongation at failure, and relaxation behavior), to encompass all potential scenarios. This analysis aims to help engineers to better predict and assess the GM behavior, specifically its ability to persist unexpected loading rates and intensities, and so significantly lower the risk of failure.

In this study, the mechanical properties are investigated through analysis of the axial true stress- axial strain relationship. An optical non-contact technique is employed to measure both axial and transverse true strains. Then, a method for determining the axial true stress is adopted.

Additionally, this research delves into the stress relaxation response associated with different tensile rates, by giving a better representation of how stress evolves over time. In this context, we consider a rheological model that is relatively simple, known as Burger's model, particularly when compared to studies that develop complex models to capture the behavior of polymeric materials (such as those by (Ghafar Chehab and Moore, 2006; Kongkitkul et al., 2014)). The proposed model combines the fundamental mechanical models of viscoelastic behavior, specifically Maxwell's and Kelvin's models (Brinson and Brinson, 2008), to depict the relaxation behavior.

## 2. Materials and Methods

### 2.1    Materials

This geomembrane considered in this work is a transparent 3 mm thick Polyvinyl Chloride geomembrane (PVC GM), obtained from a slot extrusion process through two extruders, followed by calendering technology, designed specifically for a tunnel project. The chemical composition comprises PVC resin, plasticizer, and stabilizer.

From a micro-structural point of view, PVC is made up of crystallites (called lamellae) interconnected through a three-dimensional network of polymer n-$C_2H_3$Cl. The PVC geomembrane is classified as flexible PVC, attributed to the incorporation of a plasticizer. This latter predominantly penetrates the amorphous phase, resulting in expansion in the spaces between the crystallites and, consequently, giving flexibility to the geomembrane (Summers, 1981).

The selection of this geomembrane type aims to mitigate the influence of antioxidants or any additives, such as fillers, carbon black, and fire retardants, on the research outcomes (Lodi et al., 2008). This choice allows for a more precise examination of the fundamental properties of PVC GMs, and a clearer understanding of their performance, unaffected by the existence of additives.

### 2.2    Apparatus and data acquisition

The testing procedure involves subjecting a PVC GM specimen to continuous stretching at a constant displacement rate until rupture. The experimental setup, as shown in Figure 1 comprises a tensile testing machine equipped with upper and lower grips for holding the specimen, and a force sensor with a sufficient loading capacity (at least 2000 N).

The test specimen is to be tightly clamped in the tensile test machine grips, ensuring precise alignment of the longitudinal axis of the test specimen and the axis of grips (Figure 1(a)). The grip separation speed (*i.e., displacement rate*) is adjustable within the range 0.01 to 500 mm/min. The European standard EN12311, which outlines a method for determining the tensile properties of plastic and rubber sheets, recommends a speed of 100 mm/min. The clear distance between the grips is set at 80 mm, according to the same standard.

A video extensometer is employed to measure the relative displacement between two predefined markers on the specimen. Concurrently, an acquisition system continuously records both the tensile force and the corresponding longitudinal relative displacement in the specimen. The test is conducted at a controlled temperature of 23 ± 2 °C and relative humidity in the range 30% to 70%.

## 2.3    Specimens preparation

The dimensions of the specimens for the GM uniaxial tensile test conform to the specifications mentioned in the European standard EN12311 (Method B). These specimens are randomly extracted from a roll of PVC GM. Using a template, a GM cutter machine is used to cut the specimens into a dumbbell-shaped specimen. On the less brilliant side of the specimen, two marker reference points are placed using a template, as shown in Figure 1, to measure the relative longitudinal displacement using the extensometer. The specific dimensions are detailed in Figure 1(b).

## 3. Methodology

Tensile tests, being the most straightforward experimental approach, are used for the determination of the mechanical properties of materials in terms of their stress-strain behavior, stiffness, strength, and strain at failure (*i.e., failure point where specimen is separated into two pieces*). The studied PVC GM's variation of the mechanical properties over time are investigated for short-term through strain-rate-dependent tensile tests and long-term scenarios via relaxation tests.

## 3.1 Methodology of strain-rate-dependent tensile tests

It is worth mentioning firstly that the tests conforming to the European standard EN12311 are conducted at a speed of 100 mm/min. However, this study includes a thorough evaluation of the rate dependency effect over a set of tensile speeds (*i.e., displacement rate*), spanning several orders of magnitude from 0.01 to 500 mm/min (0.01, 0.1, 0.5, 1, 10, 50, 100, 200, and 500).

Speeds exceeding 100 mm/min are included in this study to better understand and assess the material's behavior under extreme loading scenarios, such as impacts and seismic loads (Thusyanthan et al., 2012). For each tensile speed, at least three repeatability tests have been carried out. This range of displacement rates allows for a comprehensive understanding of the material's response to different loading conditions.

The test ends when the failure occurs. Throughout the test, the applied force and the corresponding relative displacement of the specimen are recorded. The outcome of this test is a force-strain curve, wherein the strain represents the engineering strain (*i.e., infinitesimal strain*) expressed as the ratio of the incremental length to the initial length (Equation 1). Such a curve founds widespread applications in industrial sectors, offering an easier and straightforward interpretation for practitioners within these fields.

$$\varepsilon_{aE} = \frac{\Delta L}{L_0} = \frac{L_f - L_0}{L_0} \#(1)$$

where $\boldsymbol{\varepsilon_{aE}}$ is the axial engineering strain or infinitesimal strain, $\boldsymbol{L_0}$ is the initial relative length between the two predefined marked points on the specimen, and $\boldsymbol{L_f}$ is the final distance between them.

Figure 2 shows the typical force-engineering strain curve. As the strain progresses, the force increases, until reaching a maximum value at the rupture point.

While force-engineering strain curves offer useful and practical insights, a deeper understanding of the mechanical properties of PVC GMs can be gained through the presentation of axial true stress – axial true strain curves (Zhang et al., 2022), providing a detailed analysis of key properties of the material, such as the maximum true tensile strength and strain. Here, axial true stress represents the axial Cauchy stress (see Equation 9), while true strain refers to the logarithmic strain.

The true axial strain is defined as the logarithmic strain, as expressed in Equation 2:

$$\varepsilon_{aT} = \int_{L_0}^{L_f} \frac{dL}{L} = \ln\left(\frac{L_f}{L_0}\right) = ln\left(\frac{L_0 + \Delta L}{L_0}\right) = \ln(1 + \varepsilon_{aE}) \#(2)$$

Both the width and the thickness of the specimen decrease during the uniaxial tensile test. Hence, similarly,the true strain for width and thickness is respectively defined as:

$$\varepsilon_{wT} = \ln(1 + \varepsilon_{wE}) \quad and \quad \varepsilon_{tT} = \ln(1 + \varepsilon_{tE}) \#(3)$$

with,

$$\varepsilon_{wE} = \frac{\Delta w}{w_0} \qquad and \qquad \varepsilon_{tE} = \frac{\Delta t}{t_0} \#(4)$$

where $\varepsilon_{wT}$ and $\varepsilon_{tT}$ are respectively the true strain for width and thickness direction, $\varepsilon_{wE}$ and $\varepsilon_{tE}$ are respectively the engineering strain in the thickness and width direction, $w_0$ and $t_0$ are respectively the initial width and thickness, $\Delta w$ and $\Delta t$ are respectively the change in width and thickness.

The relation between the axial and lateral (*i.e., width and thickness*) true strains is determined by measuring the local (i.e., logarithmic) strains along the three axes of the specimen. To do so, uniaxial tensile tests are conducted at low speed (0.5 mm/min), in order to facilitate the measurements of the thickness and width using a caliper while testing. For this purpose, two specimens, one cut along the machine direction and the second along the cross direction, are employed. In Figure 3, the relationship between the lateral true strain and the axial true strain is illustrated.

The results show a linear relationship in both the machine and cross directions up to the point of material failure. This relationship is characterized by a slope of approximately 0.5, with slight fluctuations primarily attributed to measurement errors. Therefore, the tests demonstrate that the material can be considered isotropic and incompressible.

To explore high speed conditions, two tensile tests are conducted at 100 and 200 mm/min. For these two tests, width and length are determined by a non-contact optical method using image processing with Fiji ImageJ software. The procedure of image treatment is illustrated in Figure 4. The spatial-temporal evolution of both width and length is determined by tracking pixel values across both dimensions. Remarkably, as seen in Figure 5, the results obtained from the high-speed tests reveal an axial true strain, with a slope of around 0.5, with slight deviations due to various factors such as image treatment errors, material heterogeneity, or experimental errors. Interestingly, a complementary investigation was conducted using the video extensometer analysis. The findings confirm a linear relationship between both true lateral and axial strains, with a similar slope of approximately 0.5. This consistency across testing methodologies suggests that the PVC GM under study is incompressible whatever the loading rate. Hence, the relationship between true axial and lateral strains is defined as:

$$\varepsilon_{tT} = \varepsilon_{wT} = -\alpha \varepsilon_{aT} \#(5)$$

with a value of $\alpha$ equal to 0.5.

Given the variations in cross-sectional area during testing, it becomes crucial to calculate the actual true stress, represented by the axial Cauchy stress, to accurately assess the material's behavior under applied forces. The axial Cauchy stress is determined by dividing the force acting on the material by its corresponding current cross-sectional area.

The current cross-sectional area is given by Equation 6:

$$A = w \times t \#(6)$$

where w and t denote respectively the current width and thickness, derived from Equations 3, 4 and, 5 as:

$$w = w_0 e^{-\alpha \varepsilon_a T} \qquad \#(7)$$

and,

$$t = t_0 e^{-\alpha \varepsilon_a T} \qquad \#(8)$$

where $w_0$ and $t_0$ are the initial width and thickness.

Thereupon, the axial Cauchy stress, can be calculated as follows:

$$\sigma = \frac{F}{A} = \frac{F}{w_0 t_0 e^{-2\alpha \varepsilon_a T}} \#(9)$$

Figure 6 shows the typical Cauchy (i.e., axial true) stress-logarithmic (i.e., axial true) strain curve. One can note that, until approximately 80% of axial true strain, the PVC GM behaves as a linear elastic solid. After, a nonlinear behavior prevails, underpinning an intricate interplay between elasticity, viscosity, and plasticity.

When examining the rate dependency effect, one important property to be investigated is the young modulus. Due to the non-linear stress-strain behavior of the PVC geomembrane, which lacks a defined yield point, the tensile modulus is used instead of Young's modulus.

As an estimation of the tensile modulus, the initial secant modulus is considered. This is computed by drawing a line from the origin to a specific point on the stress-strain curve, specifically at 50% logarithmic deformation, where the relationship remains approximately linear, as shown in Figure 6. To enhance accuracy, the origin point is considered at 10% logarithmic deformation to mitigate the influence of any artifacts-induced deformations caused by the initial positioning of the specimens, as shown in Figure 6.

## 3.2   Methodology of relaxation tests

Investigations through relaxation tests aim to understand the material response to persistent strain over time. To assess the effects of tensile loading rate on stress relaxation properties, seven distinct speeds have been selected: 0.01, 0.1, 0.5, 1, 10, 100, and 500 mm/min. For each speed, two tests are performed, and specimens were stretched up to a 50% axial engineering strain (to stay far away from the failure point). The decision to use engineering strain rather than true strain is driven by practical considerations arising from the testing machine, which permits the application of engineering strain but not true strain. Due to technical constraints, achieving precisely 50% of deformation was not possible; the actual deformation levels reached in each test are detailed in Table 1. Once the target strain is reached, the strain was kept constant, and the tensile stress is recorded over a duration of 4 days.

## 4.  Results and discussion

In the following, the effect of loading rates on the mechanical properties of PVC GM will be examined. In the first subsection, mechanical insights into the effect of the loading rates on the stress-strain relationships will be discussed. In the second subsection, the long-term response at constant strain will be investigated through relaxation tests. This latter study aims to provide a more comprehensive understanding of the material's response, at different loading rates, contributing to an enhanced grasp of its time-dependent behavior characteristics. Furthermore, in this section, a relatively simple viscoelastic rheological model will be developed for modeling the stress relaxation.

## 4.1   Rate-dependency response at short term

Figure 7 shows the force-engineering strain curves obtained from uniaxial tensile tests conducted at the specified aforementioned tensile rates. The recorded failure times span a range from approximately 20 seconds for the highest speed (500 mm/min) to nearly 11 days for the lowest speed (0.01 mm/min).

It is observed that for larger tensile rates the material strength increases. This observation highlights the significant effect of the tensile rate on the material's mechanical response, signaling inherent viscous properties (Meyers and Chawla, 2008).

The axial true stress- axial true strain curves for the tensile-speed-dependent tests are presented in Figure 8. It can be shown that the relationship changes with applied uniaxial tensile rate. Notably, the material becomes stiffer at higher tensile rates, while concurrently exhibiting a more malleable behavior at lower tensile rates. Further examination of this phenomenon is provided in the subsequent sections, which aim to properly analyze and elucidate the influence of applied speed on parameters such as tensile modulus and failure point (including stress and strain at failure).

### 4.1.1    Rate-dependency effect on material's stiffness

In Section 3.1, the initial secant modulus $E_{sec}$ is computed as an estimation of the tensile modulus. Figure 9 illustrates the variation of the initial secant modulus with tensile rates in logarithmic scale.

A clear dependence emerges, displaying how the material's stiffness evolves with increasing loading rates.

For further analysis, the tensile modulus (*i.e., secant modulus*) is represented against strain rate rather than tensile rate. The strain rate is determined by the ratio of the incremental length to the current length divided by incremental time, as expressed in equation 9.

$$\dot{\varepsilon}_{(t)} = \frac{\delta L}{l\,\delta t} = \frac{\Delta L}{L\,\Delta t} \#(9)$$

To illustrate, Figure 10 shows an example of the variation of strain rate during a test conducted at 1 mm/min. The strain rate reaches a peak at small deformation (i.e., at the beginning of the test), then, decreases gradually until failure occurs. In the same way, the variation of the strain rate during test is calculated for each test at each speed.

Figure 11 shows the strain rate values at the start of the test for each tensile rate. Hereinafter, the variation in tensile modulus with strain rate is presented in Figure 12, displaying a nearly linear upward trend (with the strain rate in log scale). We conjecture that it is attributed to the fact that the PVC chains, situated within a matrix composed of plasticizers, exhibit enhanced strength when subjected to high tensile rates. As chains are immersed in a bath of viscous plasticizer, an increase in the tensile rate, result in larger viscous resistance. This viscous resistance restricts the mobility of the PVC chains, inducing a stiffening effect. Consequently, the material's stiffness increases. We are currently developing a numerical model accounting for the PVC microstructure (Akel et al., 2023) that should allow us to test this conjecture.

### 4.1.2    Rate-dependency effect on the ultimate stress and strain

Examining the impact of rate dependency on the tensile strength and strain at failure is of paramount importance, as it is essential for accurately assessing the material's resistance and elongation before failure, thereby informing engineering design decisions for hydraulic projects.

Figure 13 shows the correlation between the tensile strength and strain at failure with the tensile rate. The corresponding strain rate for each speed at failure is presented in Figure 14 (to be compared to Figure 11 corresponding to the beginning of the test). It is noticeable that the failure threshold in terms of stress and strain increases with strain rate, as seen in Figure 15. Figure 13(a) reveals that by increasing tensile rate from 0.01 to 500 mm/min, corresponding to strain rates of $10^{-6}$ to $10^{-2}$ /sec (refer to Figure 15(a)), the tensile strength undergoes a considerable increase of approximately 214% (from 28 MPa to 88 MPa).  Likewise, as observed in Figure 13(b) and Figure 15(b), the strain at failure experiences a higher value with a higher loading rate, reaching 161% of axial **true strain** at the highest loading rate (500 mm/min), whereas at low loading rate (0.01 mm/min) it reaches 120% of axial **true strain**. Thus, the mechanical resistance and elongation at failure of PVC GM are strongly rate-dependent, emphasizing the importance of considering the loading rate influence in laboratory testing.

Additionally, by considering the section's surface at the failure, it has been observed that at lower tensile speeds, it displays non-smooth features and exhibits tortuosity, contrasting sharply with the entirely smooth surface observed at failure in tests conducted at high tensile speeds.

## 4.2    Rate-dependency response on the long term

Figure 16 reports the stress-relaxation curves at different tensile rates. The evolution of stress relaxation over time exhibits similar patterns across the specimens, characterized by initial quick drop followed by a progressive decrease until reaching steady regime. Depending on the tensile speeds, the curves start from different initial stress values. This initial value of stress increases with increasing loading rates, which aligns with the observation that increased loading rates lead to greater material strength, by reaching a higher value of traction forces compared to lower speeds.

### 4.2.1    Relaxation curves analysis

In order to offer a better representation of the phenomenon occurring at different time scales, Figure 17 shows the stress relaxation plotted against time on a logarithmic scale. Notably a consistent linear asymptotic tendency can be observed at the final stage of relaxation for different applied tensile rates, during the 4-day testing period.

Recently, the stress relaxation has been studied by Wu et al. (2023), showing a linear relationship between stress and time, both in logarithmic scale, for relaxation tests conducted at three specific tensile rates over only one order of magnitude: 2.5, 5, and 10 mm/min. Three initial strains have been investigated: 40.55%, 58.79%, and 81.09%. In our study, we have extended significantly the range of speeds, covering values from 0.01 to 500 mm/min. To ensure a relevant comparison with the study of Wu et al. (2023), the results are presented in a logarithmic scale for stress and strain in Figure 18. The results indicate that the relationship between ln ($\sigma$) and ln ($t$) is not consistently linear. Particularly, as the tensile rate reduces, the curve tends to deviate, and a more horizontal trend emerges at the beginning stage of the relaxation curves for low tensile rates.

As noted by Wu et al. (2023), the initial strain significantly affects the relaxation stress. However, the specimens have not achieved the same strain level, as reported in Table 1. Thus, this may introduce some variability in the stress values at the onset of the relaxation. Therefore, these relaxation curves require normalization to get rid of the impact of the deformation level at the beginning of the relaxation phase. In that way, this step ensures that the variations in the initial deformation do not affect the interpretation of stress-relaxation behavior.

In this normalization method, the asymptotic line for each test is determined. This line represents the long-term behavior of the stress-relaxation, typically modeled as:

$$\sigma = \sigma_{asymptote} + b \ln \left( \frac{t}{t_0} \right) \#(10)$$

where $\sigma_{asymptote}$ is the asymptotic stress at $t_0 = 1 \text{min}$, b is the slope of the asymptotic line, and $t$ is the elapsed time in minutes.

As seen in Figure 19, both the asymptotic stress and the slope of the asymptotic line are affected. This is due not only to the initial strain level but also to several additional factors such as material heterogeneity and experimental errors. The normalization step consists in reducing the effects of the initial strain as much as possible while neglecting other factors. This allows for a more direct comparison of the initial and intermediate stress relaxation stages across tests performed at different tensile rates, in order to assess their impacts on the relaxation behavior. The method consists in subtracting the asymptotic stress ($\sigma_{asymptote}$) from the observed stress values for each test, as follows:

$$\sigma_{(t)}^{corrected} = \sigma_{(t)} - \sigma_{asymptote} \#(11)$$

As observed in the normalized relaxation curves in Figure 20, there are two distinct relaxation time characteristics. The first one is highly dependent on the applied tensile loading rate prior to relaxation, while the second one, marked by a consistent linear tendency in the final stage of relaxation, seems not to depend on the applied tensile loading rate. This suggests that, after a certain time – approximately 240 minutes based on graphical analysis – the material exhibits the same behavior across all specimens, indicating the occurrence of a common material phenomenon. It is worth to mention that this time may vary according to environmental conditions, such as temperature and (to a smaller extent) humidity.

The behavior, shown within the initial four hours of relaxation, might be attributed primarily to the plasticizer. During this phase, we hypothesize that the resistance forces generated during the pre-relaxation phase (i.e., tensile stretching to a defined deformation level) gradually dissipate, leaving the polymer chains to govern the relaxation behavior. As long as the stress within the chains remains below a certain threshold strength, the material exhibit longer time resistance. The gradual decrease in stress over time during this phase might be attributed to the thermal fluctuations of the polymer chains occurring at the molecular level.

### 4.2.2 Rheological analysis

Considerable research has been dedicated to formulating a constitutive model for theoretically capturing the mechanical behavior of Polymeric material (Ghafar Chehab and Moore, 2006; Kongkitkul et al., 2014; Kuhl et al., 2016; Yin et al., 2019; Zhang and Moore, 1997). In 1D conditions, the simplest mechanical models for viscoelastic behavior consist of two elements: a spring for the elastic behavior and a damper for the viscous behavior. Among these models, Maxwell's model, consisting of a linear spring and a linear viscous damper in series, has proven to be effective in describing relaxation behavior. In contrast, Kelvin's model, featuring these components in parallel, is less accurate for stress relaxation but more suitable for creep behavior (Brinson and Brinson, 2008). These aforementioned models represent the backbone of various advanced models used in describing the mechanical behavior of viscoelastic and viscoplastic materials. Therefore, the combination of spring and damper elements in various configurations allows for simulating diverse viscoelastic responses. In relaxation, models incorporating a free damper result in stress decay to zero, whereas models without a free damper lead to stress decay approaching a non-zero limit value.

We propose to use Burger's model, with two dashpot dampers, in order to account for the double-time characteristics, as illustrated in Figure 20. This extended model consists in one-kelvin sub-model with a spring and dashpots in series, as illustrated in Figure 21.

The generalized constitutive equation is determined as follows:

$$\sigma + \left[\frac{\eta_1}{K_{v1}} + \frac{\eta_2}{K_{v1}} + \frac{\eta_2}{K_e}\right]\dot{\sigma} + \frac{\eta_1\eta_2}{K_e\,K_{v1}}\,\ddot{\sigma} = \eta_2\dot{\varepsilon} + \left(\frac{\eta_1\,\eta_2}{K_{v1}}\right)\ddot{\varepsilon} \# (12)$$

In relaxation regime, where $\varepsilon$ is constant, $\dot{\varepsilon}$ and $\ddot{\varepsilon}$ are both zero. Consequently, the equation describing relaxation simplifies as:

$$\sigma_t = A\,e^{-\frac{1}{\tau_1}t} + B\,e^{-\frac{1}{\tau_2}t} \# (13)$$

with:

$$\tau_1 = \frac{2\eta_1\eta_2}{\eta_1 K_e + \eta_2(K_{v1} + K_e) + \sqrt{[\eta_1 K_e - \eta_2(K_{v1} + K_e)]^2 + 4\eta_1\eta_2 K_e^2}}$$

$$\tau_2 = \frac{2\eta_1\eta_2}{\eta_1 K_e + \eta_2(K_{v1} + K_e) - \sqrt{[\eta_1 K_e - \eta_2(K_{v1} + K_e)]^2 + 4\eta_1\eta_2 K_e^2}}$$

where $A$ and $B$ are constants that depend on the applied tensile rate prior to relaxation, $t$ is the time.

The results indicate a strong correlation with the 0.01 mm/min test speed, evidenced by an $R^2$ value of 0.98 (Table 2). The two characteristics times $\tau_1$ and $\tau_2$ are respectively 209.86 min $\simeq$ 3.5 h and 23,691 min $\simeq$ 16.45 days, where $\tau_1$ is close to the threshold time 240 min that separate visually the two regimes in the relaxation curve. Nevertheless, as the speed increases, Burger's model becomes less accurate to depict the entire relaxation behavior (Figure 22).

Therefore, the proposed rheological model capture well the existence of two distinct characteristic times, but it does not capture the linear decrease of log (σ) with respect to log (t) on the longer term.

Consequently, the usefulness of the model is not to provide accurate results for long-term relaxation but to put forward the existence of two distinct relaxation mechanisms. The understanding of the microscale physics behind needs further investigations.

## 5. Conclusions

In this study, the time dependent response of PVC GM is investigated. The short-term investigation through strain-rate-dependent tensile tests is carried out by conducting uniaxial tensile tests with applying constant displacement rates. This work offers an original analysis that considers a broad set of tensile loading rates with varying order of magnitude. The results reveal a significant impact of applied tensile speed on stress-strain curves, indicating that higher loading rates are associated with an increase in tensile modulus. Moreover, mechanical properties at failure exhibit time-dependency, displaying a reduction in strength and strain at failure as loading rates decrease.

Exploring the effects of loading speed on material behavior, in this study, not only sheds light on fundamental aspects but also provides valuable insights for engineering applications. This study demonstrates the impact of the tensile rate on the mechanical resistance and deformation at failure. Hence, for an engineering safety standpoint, the relatively short-term laboratory testing for predicting the resistance of PVC's GM should be performed at the lower rates expected in the field, in order to accurately measure intrinsic ultimate strength and deformation, and to avoid the risk of premature failure. Relying solely on tests at 100 mm/min, the standard laboratory test - as recommended by European standard

EN12311 - may not be conservative, as failure occurred sooner for lower speeds. However, based on this study, it remains unclear whether the lowest speed of 0.01 mm/min would result in ultimate intrinsic strength and deformation. Therefore, additional tests at significantly lower speeds, potentially decreasing by a factor of 10, are needed to comprehensively assess the material's ultimate intrinsic strength and strain.

Long-term effect of time have been also investigated through relaxation tests. To evaluate the impact of tensile speeds on stress relaxation properties, a comprehensive set of seven distinct tensile speeds ranging from 0.01 to 500 mm/min is chosen. These tests are conducted at a predefined initial engineering strain of 50%. According to the stress relaxation curves, the results demonstrates that the relaxation is classified into two typical stages, characterized by two distinct time characteristics. The dividing point between the two stages, at 23°C, is approximately located at a relaxation time of 240 min (4h). The first stage depends highly on previous loading rate history. In contrast, the second stage demonstrates independence from past loading rate, highlighting that, at the long term, pre-relaxation loading rate will not significantly influence the mechanical properties, and same phenomena occur at the microstructure level producing the same observed mechanical response. Furthermore, in this study, rheological model has been developed to predict relaxation behavior. Two dashpot dampers are incorporated in order to validate the presence of to two distinct time characteristics. This model exhibits satisfactory fitting at low loading rates (0.01 mm/min), confirming the existence of two characteristic times associated with distinct phenomena. However, beyond this tensile speed, the model with constant parameters is unable to accurately predict the relaxation behavior, requesting variable parameters.

Consequently, this study proves that in long-term testing of material behavior, such as relaxation behavior presented in this study, the loading speed no longer significantly affects measurements after few hours (approximately 4 hours). This suggests that for extended testing periods, such as in a creep test, the tensile speed testing will not affect the results at long term.

This consideration may guide experimental decisions when investigated the behavior during prolonged deformation testing, such as in creep studies.

We have proposed to interpret this rate-dependency behavior from the interplay between polymers and plasticizer, where each of which bears a part of the stress. We conjecture that at higher loading rates, the increased resistance originates from the viscous behavior of the plasticizer where it restricts the mobility of the PVC chains, inducing an increase in material stiffness. Our ongoing development of a 3D micromechanical model (Akel et al., 2023) aims to test this hypothesis in future research.

The results discussed in this paper are specific to the geomembrane being studied and should be extended carefully to other types of geomembranes in which other additives (such as antioxidants, fillers, carbon black, and fire retardants) may influence the mechanical behavior as well. Considering the distinctive microstructure of each geomembrane type, its individual response to tensile forces and long-term performance may vary accordingly.

## 6. Acknowledgments

The authors acknowledge the financial support provided by the National Research Institute for Agriculture, Food, and the Environment (INRAE) and the French Région Sud Provence-Alpes-Côte d'Azur (PACA). Additional support is gratefully acknowledged from the operating budget provided by the industrial partner EGC Galopin. The writers are also grateful to Julien Aubriet, Naïm Chaouch and Abbas Farhat for their assistance in conducting the experimental tests.

## 8. List of Figures

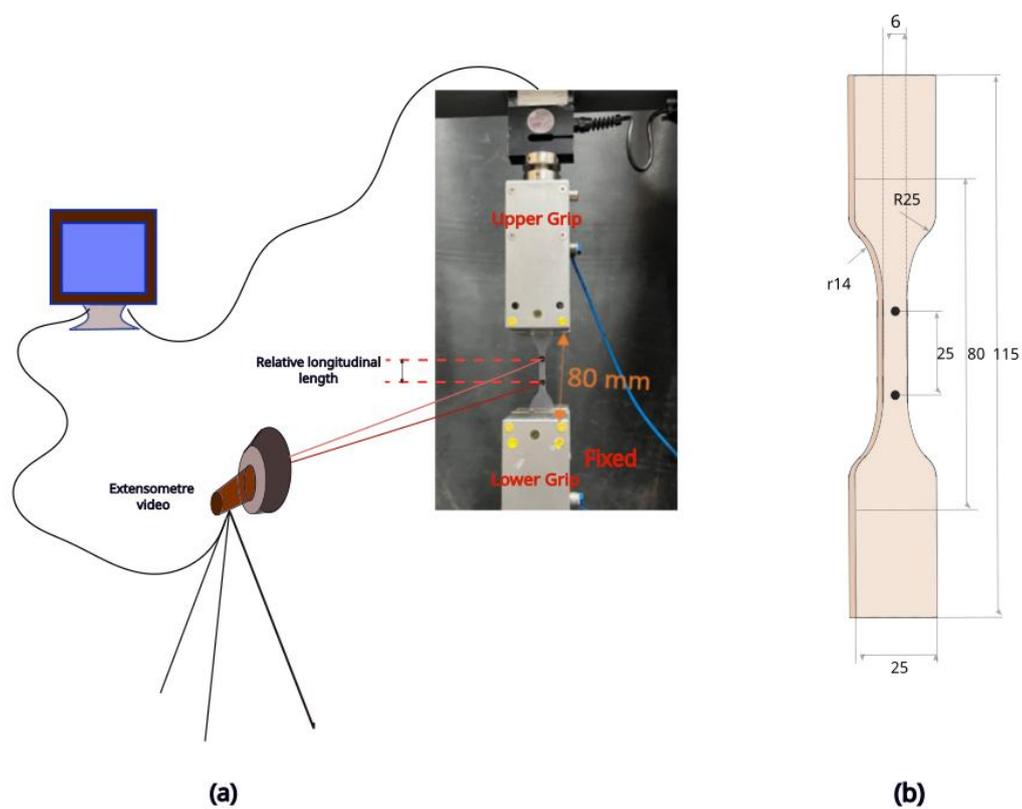

**(a)**             **(b)**

Figure 1: (a) Specimen mounted between the clumps of the two grips. (b) Dimensions of Dumbbell-shaped specimen

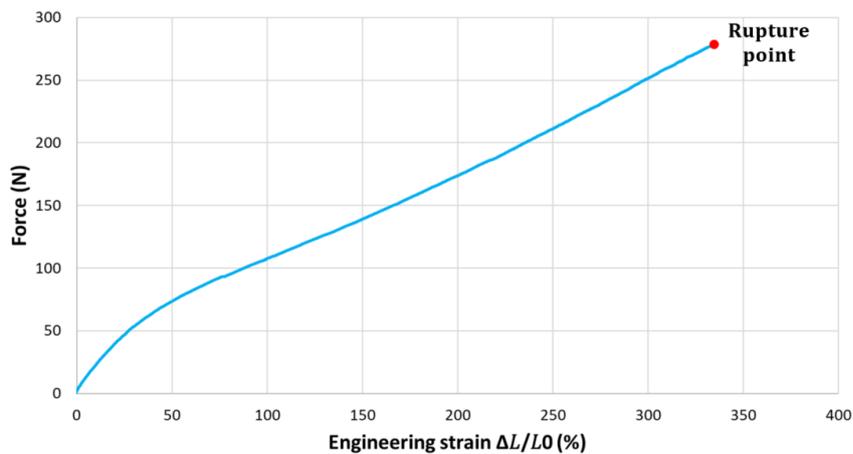

Figure 2: Typical force-engineering strain curve in uniaxial tensile tests.

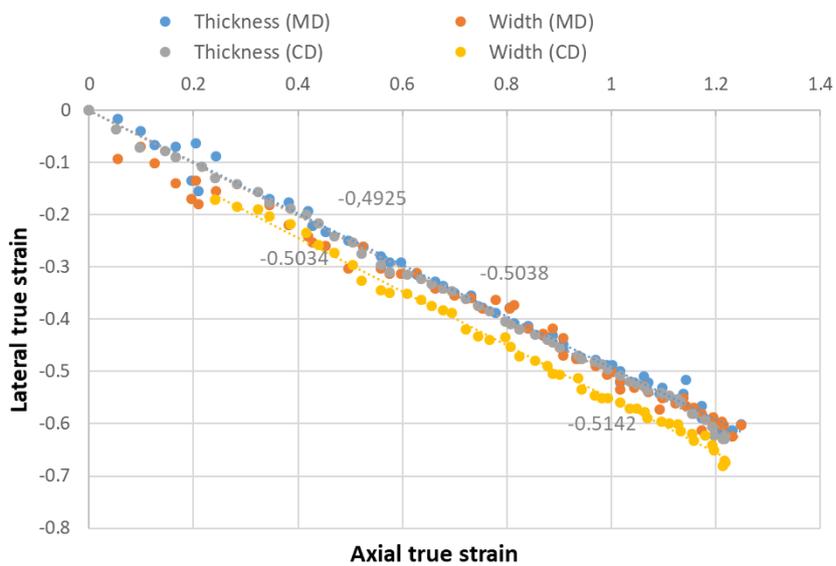

Figure 3: Lateral true strain vs. axial true strain in machine (MD) and cross directions (CD) at a uniaxial

tensile speed of 0.5mm/min:  Slopes of -0.49 (Thickness) and -0.5 (Width) in MD, and -0.5 (Thickness) and -0.51

(Width) in CD.

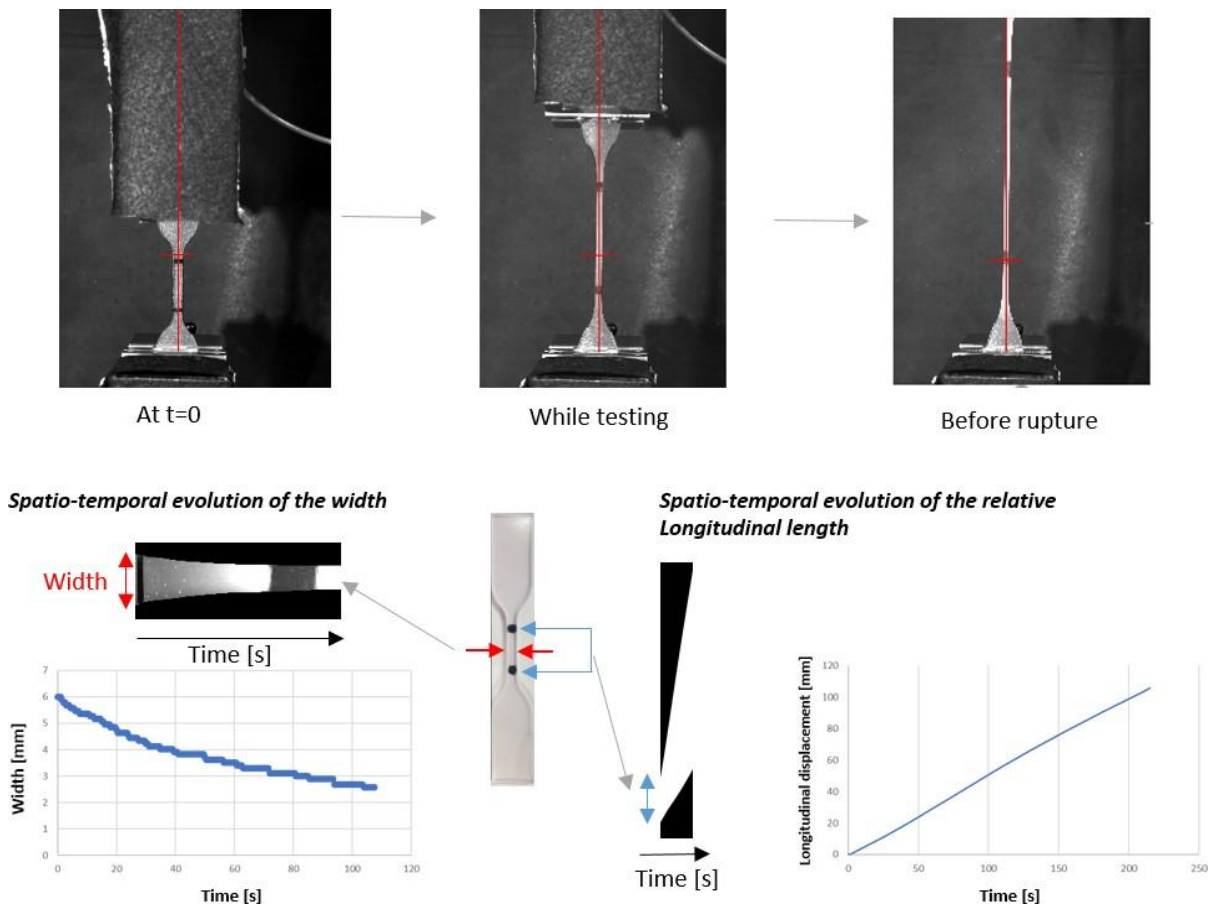

Figure 4: Schematic view of the image processing method employed to determine the width and relative longitudinal length during uniaxial tensile testing. The spatial-temporal evolution of both width and length is determined by monitoring pixel values along the width and length using a horizontal and vertical red line, respectively.

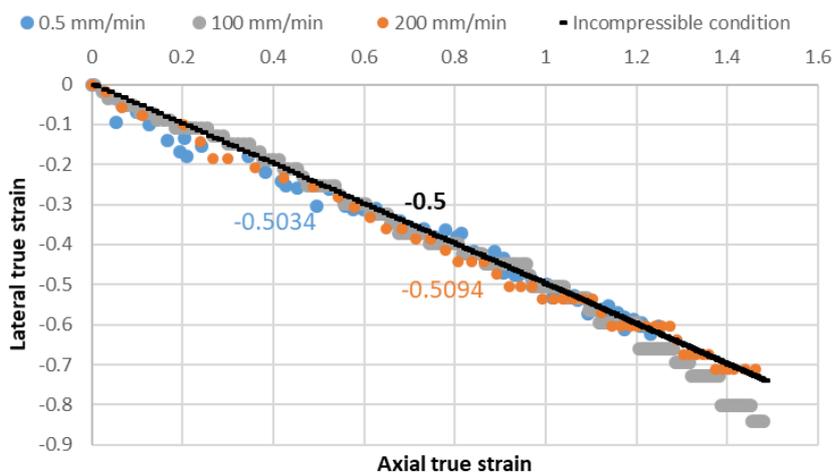

Figure 5: Lateral true strain vs. axial true strain at uniaxial tensile speeds of 0.5, 100, and 200 mm/min.

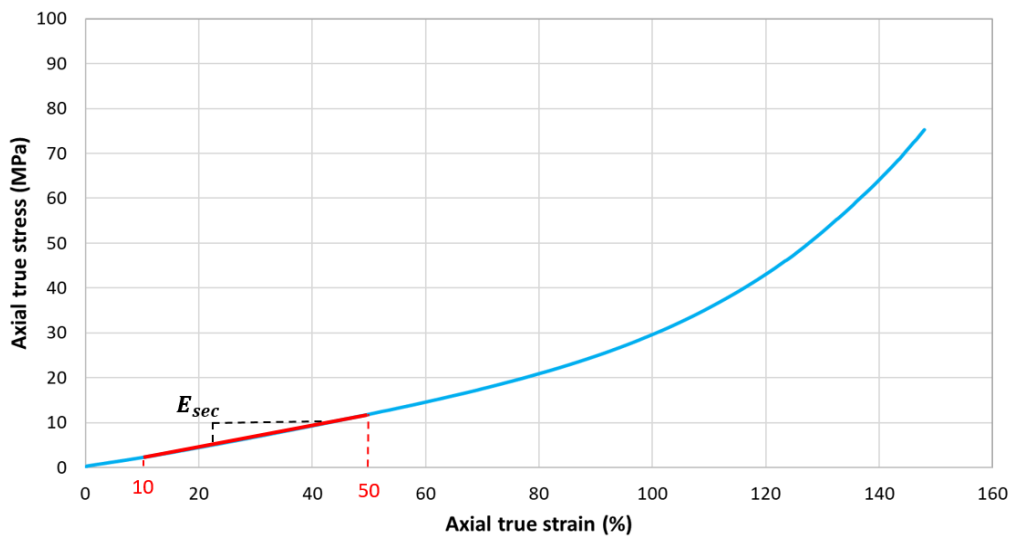

Figure 6: Typical axial true stress - axial true strain curves in uniaxial tensile tests. The tensile modulus $E_{sec}$

is determined from the initial secant drawn between 10 % and 50 % true strain.

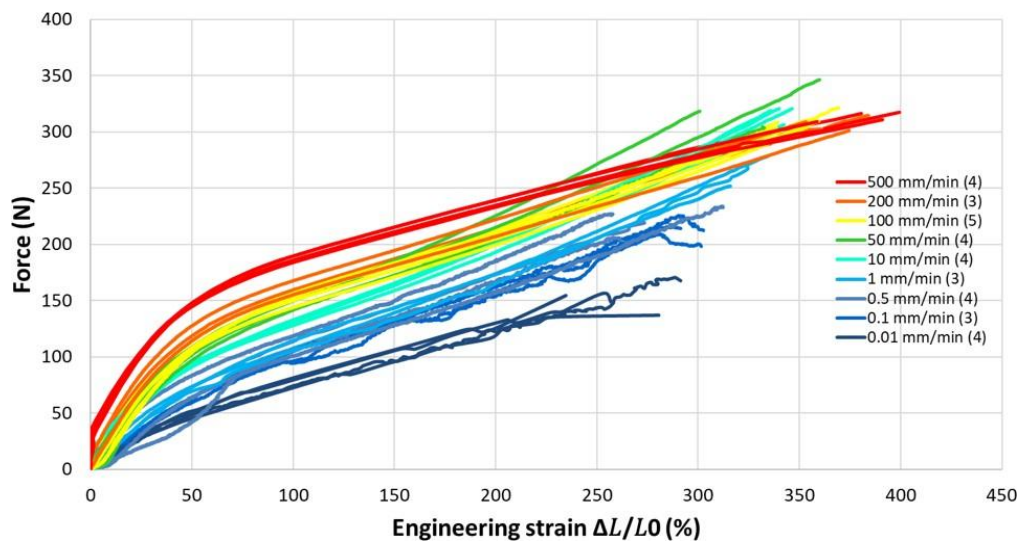

Figure 7: Force-engineering strain curves obtained from uniaxial tensile tests conducted at different tensile

rates (mm/min). The values between brackets in the legend represent the number of repeatability tests

conducted for each tensile speed. The standard laboratory test, as recommended by European standard

EN12311, involves conducting a tensile test at a prescribed uniaxial displacement rate of 100 mm/min

(corresponding to yellow curves).

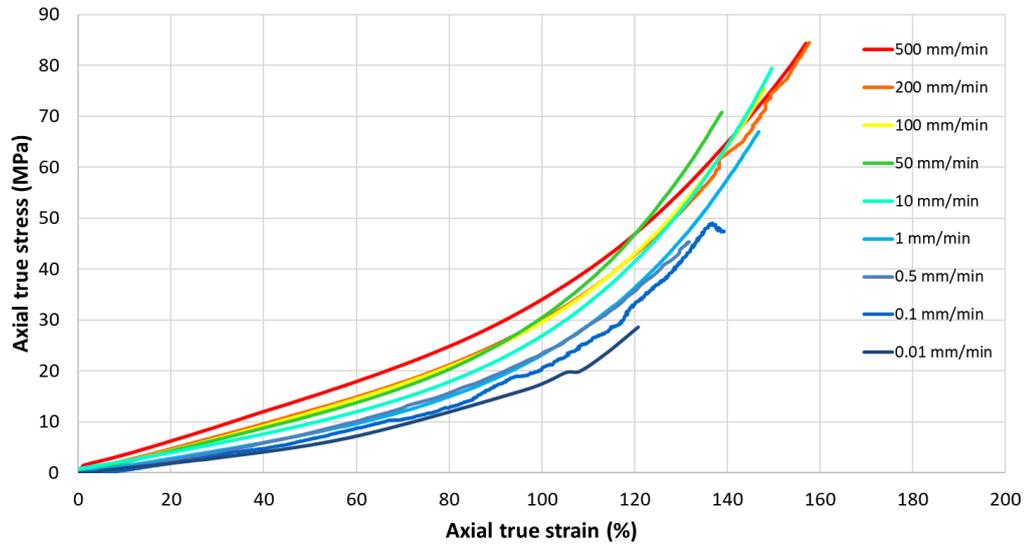

Figure 8: Stress-strain curves for strain-rate-dependency tests. The curves exhibit a distinctive shape compared to the force-deformation curve, which is due to the reduction in cross-sectional area. To improve clarity of the observations, only the data for Specimen 1 is presented for each tensile rate.

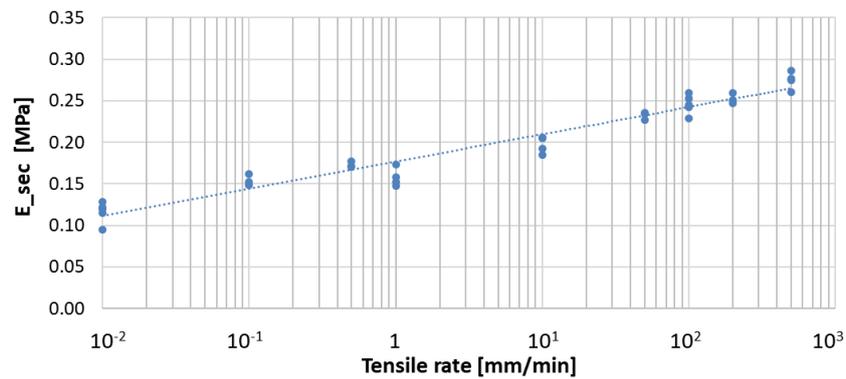

Figure 9: Variation of tensile modulus with tensile rate.

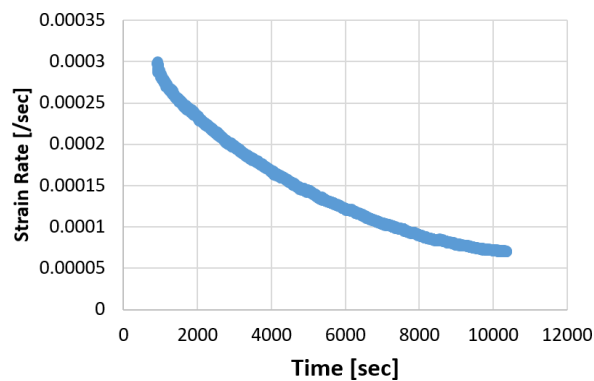

Figure 10: Variation of the strain rate over time during a uniaxial tensile test conducted at 1 mm/min.

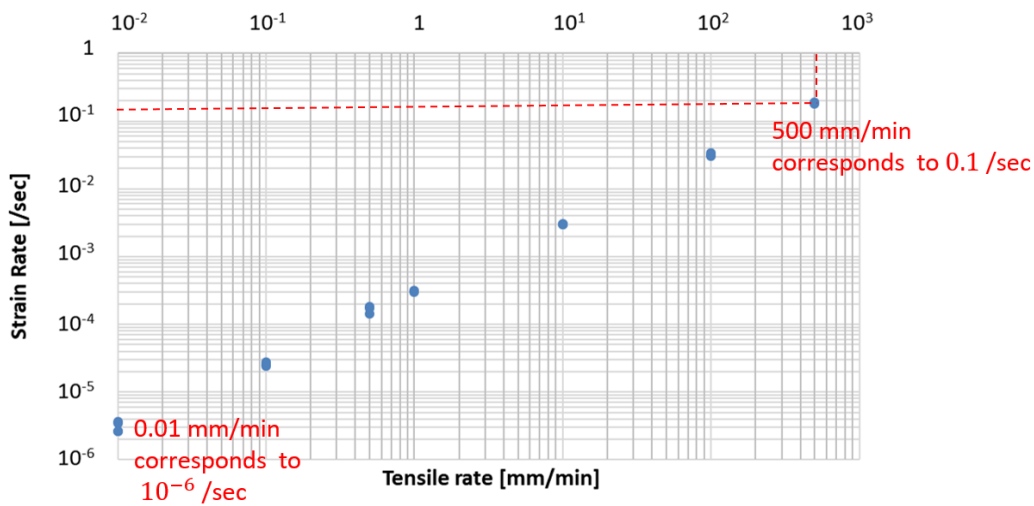

Figure 11: Strain rate at the beginning of the test for different tensile rates.

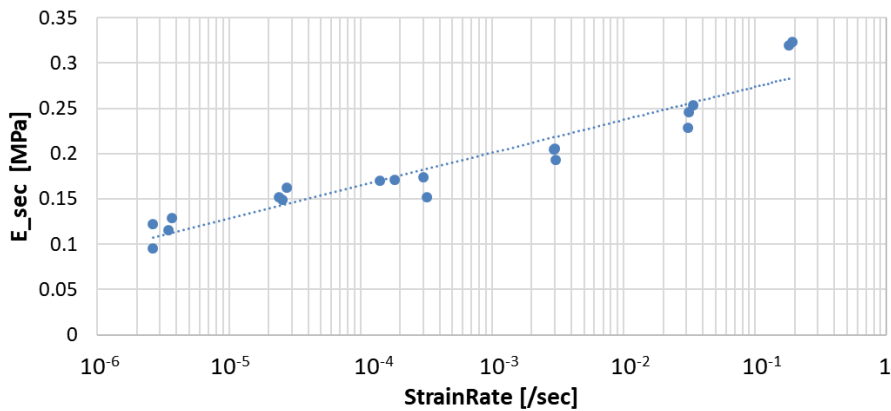

Figure 12: Variation in tensile modulus with strain rate.

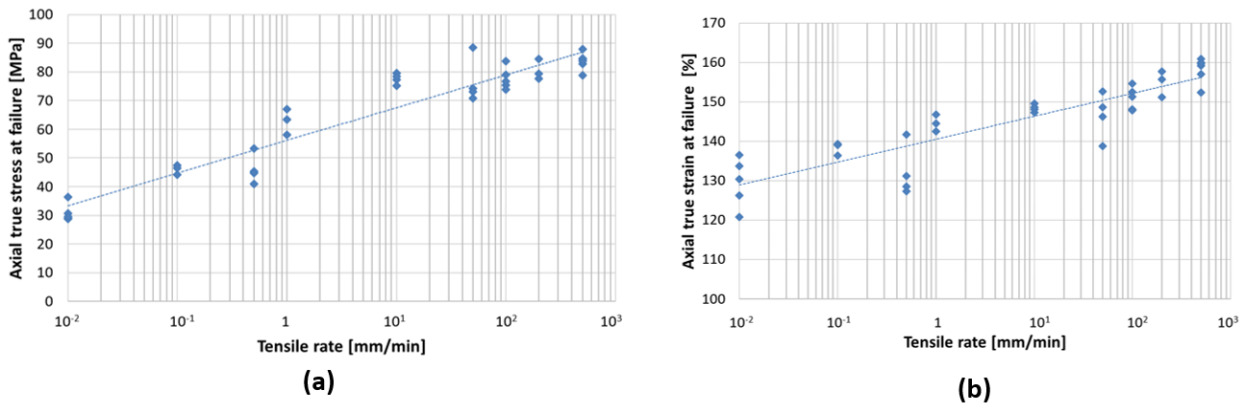

Figure 13: Variation of axial true stress and axial true strain at failure with tensile speed.

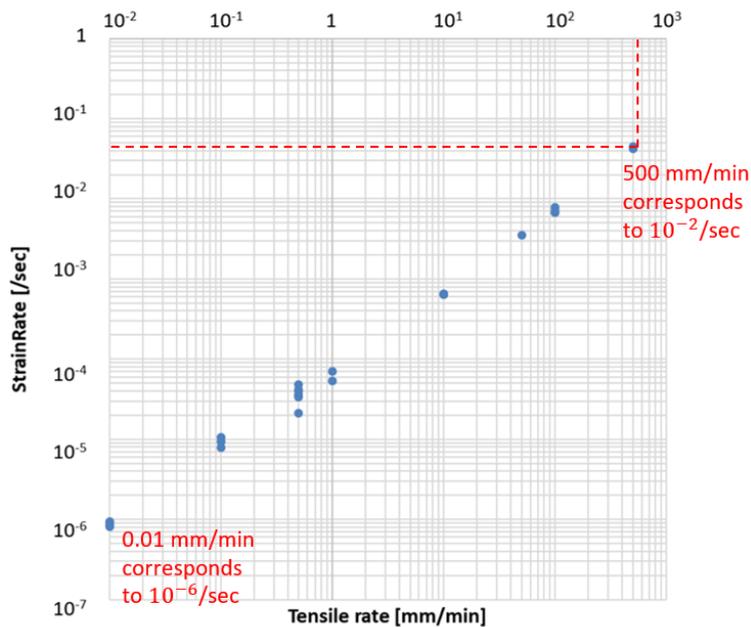

Figure 14: Strain rate at failure for each tensile speed.

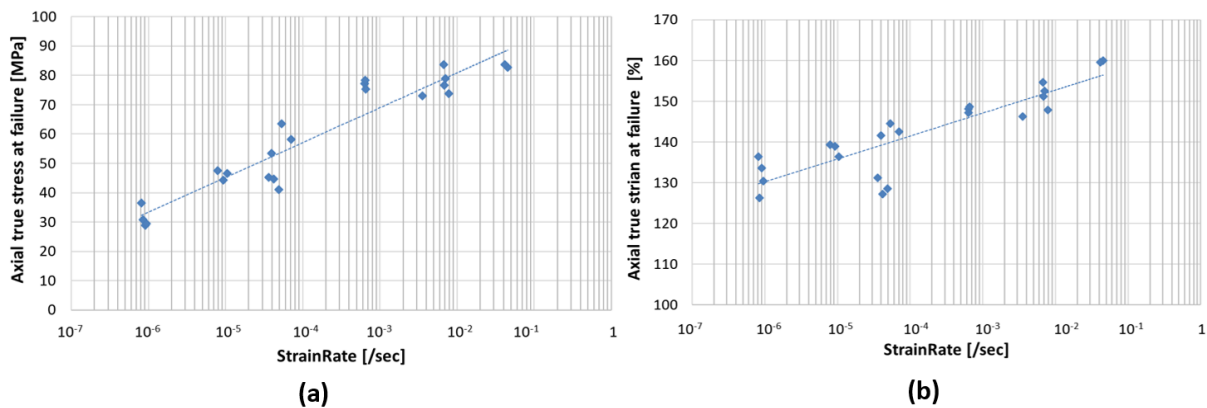

**(a)**          **(b)**

Figure 15: Variation of axial true stress and axial true strain at failure with strain rate.

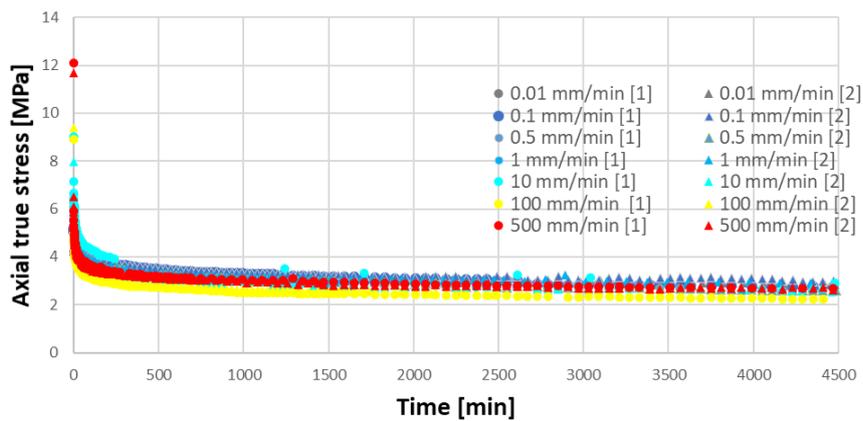

Figure 16: Stress relaxation curves at different tensile speeds.

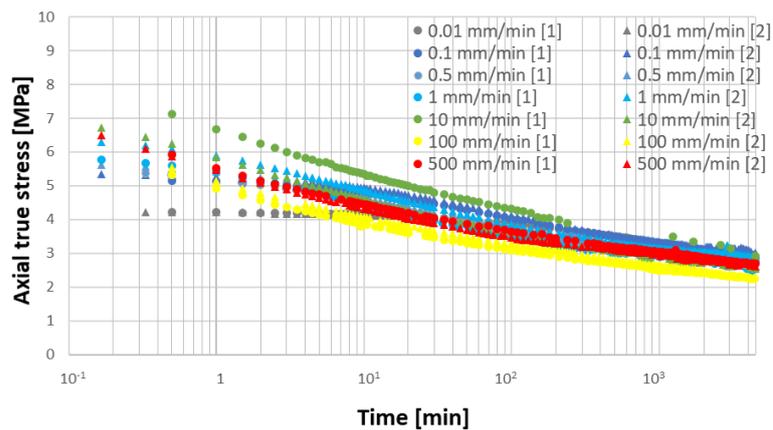

Figure 17: Stress relaxation curves at different tensile rates in logarithmic time scale.

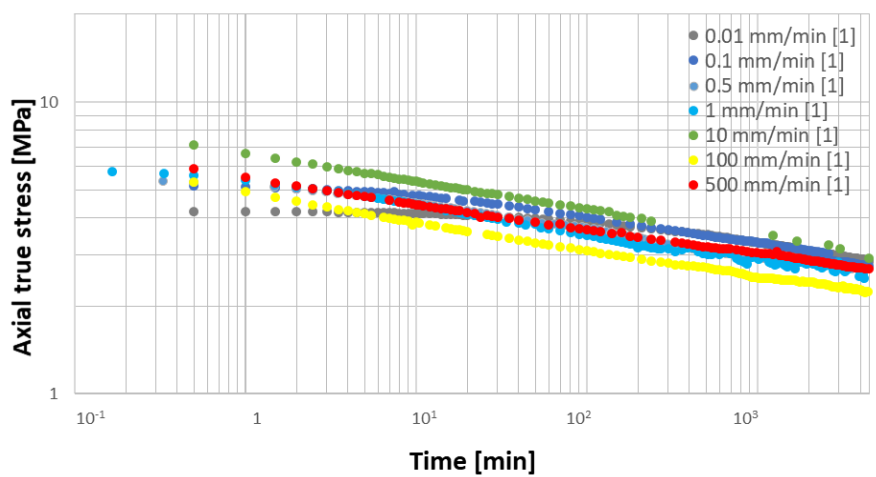

Figure 18: Stress relaxation curves presented in logarithmic scales for both axes at various tensile

rates. To improve clarity of the observations, only the data for Specimen 1 is presented.

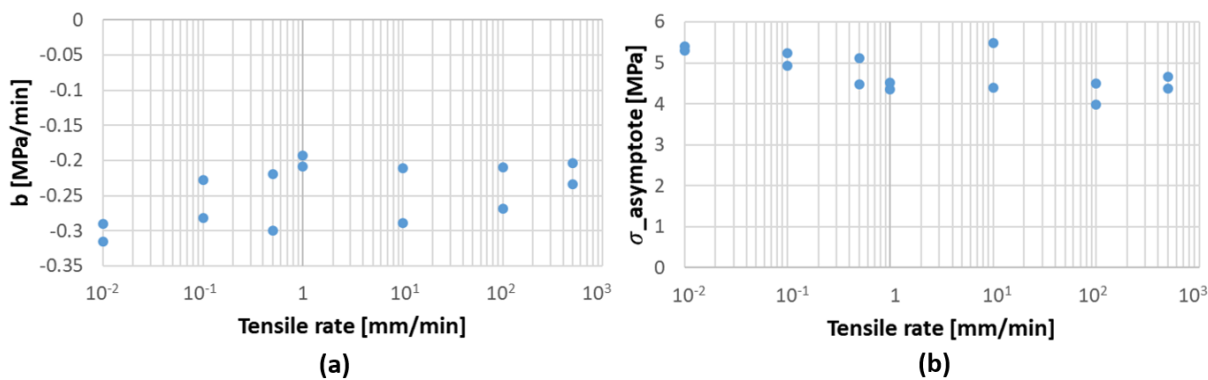

Figure 19: Variations of: (a) the slope b of the asymptotic line for each test, (b) $\sigma_{asymptote}$.

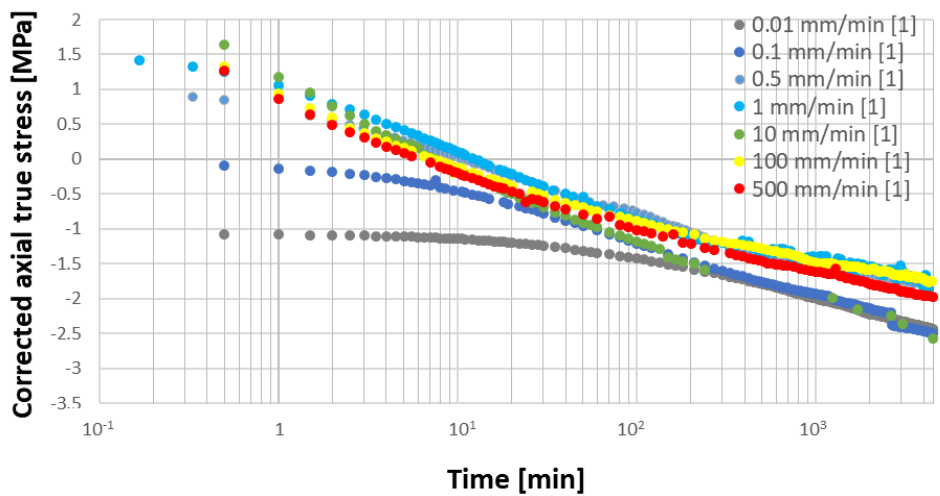

Figure 20: Corrected stress relaxation over time. To improve clarity of the observations, only the data for

Specimen 1 is presented.

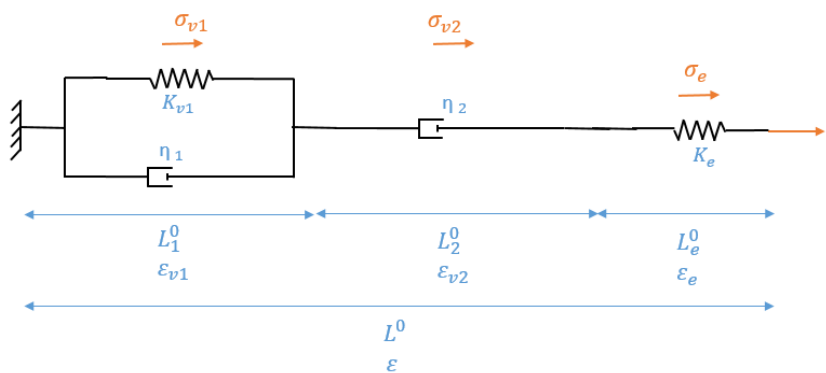

Figure 21: Extended, five-parameter Burger's rheological model.

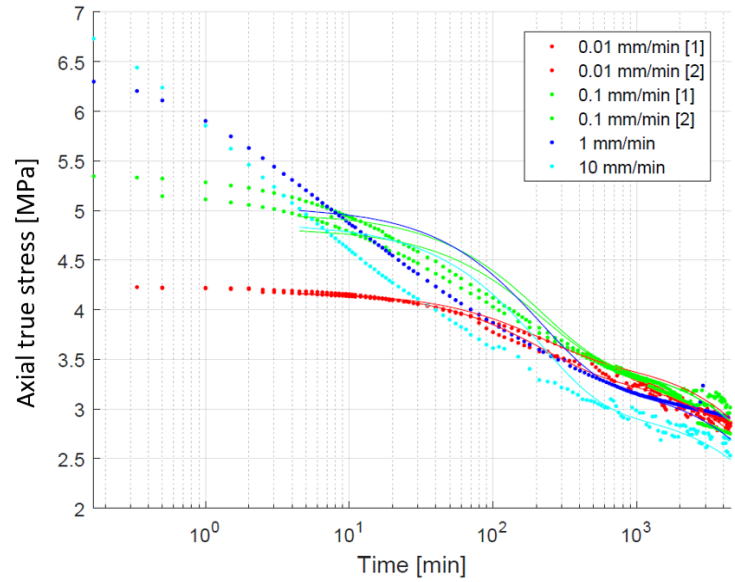

Figure 22: Fitting results of the relaxation curves at 0.01, 0.1, 1 and 10 mm/min.

## 9. List of Tables

Table 1: Level of maximal engineering strain achieved during each relaxation test.

| Tensile speed [mm/min] | Level of engineering strain[%] | |
|---|---|---|
| | Test 1 | Test 2 |
| **0.01** | 57.44 | 50.64 |
| **0.1** | 49.35 | 53.31 |
| **0.5** | 46.62 | 49.92 |
| **1** | 49.97 | 49.97 |
| **10** | 61 | 50.06 |
| **100** | 50.09 | 50.22 |
| **500** | 50.17 | 49.15 |

Table 2: Model parameters determined from test data fitting.

| Tensile loading rate [mm/min] | A | B | $R^2$ |
|---|---|---|---|
| **0.01** | 0.6696 | 3.507 | 0.9078 |
| **0.01** | 0.8305 | 3.348 | 0.9889 |
| **0.1** | 1.346 | 3.477 | 0.9092 |
| **0.1** | 1.499 | 3.475 | 0.9663 |
| **1** | 1.779 | 3.258 | 0.8707 |
| **10** | 1.863 | 3.008 | 0.8135 |